\documentclass[twocolumn,twoside,slac_two]{revtex4}
\usepackage{graphicx}
\usepackage{fancyhdr}
\usepackage{amsmath}
\usepackage{amssymb}
\usepackage{units}
\usepackage{booktabs}

\begin{document}

\title{Search for transient neutrino sources with IceCube}

%

\author{A. Franckowiak for the IceCube Collaboration}
\affiliation{Physikalisches Institut, Universit\"{a}t Bonn, Nussallee 12, D-53113 Bonn, Germany}

\begin{abstract}
The IceCube detector, which is embedded in the glacial ice at the geographic South Pole, is the first neutrino telescope to comprise a volume of one cubic kilometer.
The search for neutrinos of astrophysical origin is among the primary goals of IceCube. Point source candidates include Galactic objects such as supernova remnants (SNRs) as well as extragalactic objects such as Active Galactic Nuclei (AGN) and Gamma-Ray Bursts (GRBs).
Offline and online searches for transient sources like GRBs and supernovae (SNe) are presented. Triggered searches use satellite measurements from Fermi, SWIFT and Konus. Complementary to the triggered offline search, an online neutrino multiplet selection allows IceCube to trigger a network of optical telescopes, which can then identify a possible electromagnetic counterpart. This allows to probe for mildly relativistic jets in SNe and hence to reveal the connection between GRBs, SNe and relativistic jets.
Results from IceCube's triggered GRB search and a first limit on relativistic jets in SNe from the optical follow-up program are presented.
\end{abstract}

\maketitle

\thispagestyle{fancy}


\section{Introduction}
Gamma-ray Bursts (GRBs) are prime candidates for the production of the highest energy cosmic rays because of the enormous energy that is released in such an event \citep{waxman95} ($\mathcal{O}( 10^{51} - 10^{54} \unit{erg} \times \Omega / 4 \pi)$ in $\gamma$-rays, where $\Omega$ is the opening angle of a possible beamed emission). If the prime engine accelerates protons and electrons with similar efficiencies this would be sufficient energy to account for the observed ultra high energy cosmic rays. The observed $\gamma$-rays would originate from high energy electron synchrotron emission and inverse Compton scattering, while high energy neutrons would escape the fireball's magnetic field and later decay to protons, which would be responsible for the high energy cosmic ray flux seen on Earth. The observation of keV to MeV $\gamma$-rays confirms the presence of high energy electrons in the fireball; however, because high energy protons are deflected in inter-galactic and the Galactic magnetic fields no direct observation of protons from GRBs is possible. Nevertheless, if high energy protons are present in the fireball along with high energy electrons it is reasonable to assume that pions will be produced through interactions of protons with $\gamma$-rays from electron synchrotron radiation near the source, which would give rise to neutrinos. \cite{guetta2004} give a detailed account of the expected neutrino flux from such interactions. Here we present the most recent search for neutrinos in temporal and directional coincidence with reported GRBs done in IceCube.\\ 
Recent observations imply a common physical origin of long GRBs and core-collapse supernovae (CCSNe): a massive stellar explosion (see~\cite{Woosley:2006fn} for a review). 
According to the collapsar model \citep{MacFadyen:1998vz}, long GRBs (duration $\gtrsim 2$\,s) have their origin in 
the collapse of a massive, rapidly rotating star into a black hole surrounded by an accretion disk. Relativistic jets 
with Lorentz boost factors of 100-1000 form along the stellar axis.
This GRB-SN connection gives rise to the idea that GRBs and SNe 
might have the jet signature in common and a certain fraction of core-collapse SNe host soft relativistic jets. 
SN jets are suggested to be equally energetic and more baryon-rich, hence they are only mildly relativistic.
Such soft relativistic jets would become stalled in the outer layers of the progenitor star, leading to essentially full absorption of
the electromagnetic radiation emitted by the jet and at the same time an efficient production
of high-energy neutrinos \citep{Razzaque:2005bh,AndoBeacom}. This motivates a search for neutrino emission from SNe, as neutrinos would be able to escape from within the star. \cite{AndoBeacom} present a detailed calculation of the expected neutrino flux. \\
A dedicated search for a neutrino signal in coincidence with the observed X-ray flash of SN~2008D has been conducted by IceCube \citep{Abbasi:2011wh} in order to test the soft jet scenario for CCSNe. 
Early SN detections, as in the case of SN~2008D, are very rare since X-ray telescopes have a limited field of view. However, neutrino telescopes cover half of the sky with high sensitivity at any time. 
If neutrinos produced in soft relativistic SN jets are detected in real time, they can be used to trigger follow-up observations \citep{Kowalski:2007xb}. This is realized with an online analysis of neutrino data combined with an optical follow-up.
This optical follow-up program is independent of satellite detections and thus complementary to the triggered searches mentioned above. It is sensitive to transient objects, which are either $\gamma$-dark or
missed by $\gamma$-ray satellites. In addition to a gain in significance, the optical observations may allow
to identify the transient neutrino source, be it a SN, GRB or any other 
transient phenomenon producing an optical signal. Hence it enables us to test 
the plausible hypothesis of a soft relativistic SN jet
and sheds light on the connection between GRBs, SNe and relativistic jets.\\
While up to now the triggered search is performed offline on an entire dataset ($\sim1$ year of data) with time consuming
reconstructions on a large computer cluster, the optical telescopes have to be triggered in real time and thus require an online analysis of the neutrino data.

\section{The IceCube Detector}
\label{sec:NeutrinoDetection}
The IceCube neutrino telescope \citep{IceCube} has been under construction at the geographic South Pole 
since 2004 and was completed in the Antarctic summer of 2010/11. It is capable of detecting high energy neutrinos with energies of $\mathcal{O}$(100)GeV and is most sensitive to muon neutrinos with energies in the TeV range and above. 
High-energy muon neutrinos undergoing charged current interactions in the ice or the underlying rock
produce muons in neutrino-nucleon interactions.
The muon travels in a direction close to that of the neutrino and emits Cherenkov light. 
The deep ultra clear Antarctic ice is instrumented with light sensors thus forming a Cherenkov particle detector. After its completion it comprises a volume of $1$\,km$^3$ with $5160$ digital optical modules (DOMs) attached to $86$ vertical strings at a depth of $1450$\,m to $2450$\,m. Each DOM consists of a $25$\,cm diameter Hamamatsu 
photomultiplier tube (PMT) and supporting hardware inside a 
glass pressure sphere. Here we present the offline analysis of data taken with the 40-string detector configuration (2008/04/05 to 2009/05/20) and the 59-string configuration (2009/05/20 to 2010/05/25) as well as the online analysis of the data taken from the start of the follow-up program on 2008/12/16 to 
2009/12/31.
In the following the deployment stages will be referred to as IC40 and IC59.

\section{Triggered Offline Analysis}

The offline analysis is triggered by satellite detections. It looks for neutrinos coincident in space and time with the observed $\gamma$-radiation.

\subsection{ The GRB sample }
During the IC59 data taking period, 105 GRBs were observed in the northern sky and reported via the GRB Coordinates Network (GCN\footnote{\url{http://gcn.gsfc.nasa.gov}}). Of those GRBs 9 had to be removed, because IceCube was not taking physics data. 
The GRB localization is taken from the satellite that has the smallest reported error. The start ($\mbox{T}_{\mbox{start}}$) and stop ($\mbox{T}_{\mbox{stop}}$) times are taken by finding the earliest and latest time reported for $\gamma$-emission. The fluence, and $\gamma$-ray spectral parameters are taken preferentially from Fermi (GBM), Konus-Wind, Suzaku WAM, and \textit{Swift} in this order.
The $\gamma$-ray spectra reported by the satellites were used to calculate the neutrino spectra and flux as outlined in Appendix A of \cite{guetta2004}. 
GCN does not always report values for all of the parameters used in the neutrino spectrum calculation. In that case average values from \cite{Abbasi:2011qc} are used for the parameters not measured by the satellites. 

\begin{figure}[t]
  \centering
  \includegraphics[width=\linewidth]{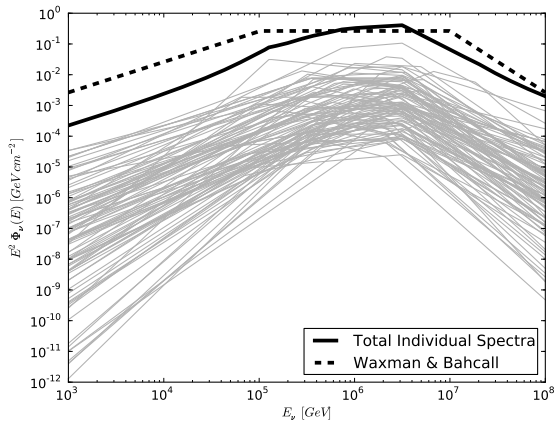}
  \caption{Neutrino spectra of the GRBs used in the triggered offline analysis. The thin lines represent the individual bursts while the solid thick line represents the sum of all bursts. Finally, the dashed line shows the \cite{Waxman2003} prediction normalized to the number of GRBs observed.}
  \label{fig:neutrino_spectra}
\end{figure}

\subsection{ Offline Event Selection }
The analysis presented here was designed to be sensitive to neutrino production from $p\gamma$ interactions in the prompt phase of the fireball. To separate signal from background a Boosted Decision Tree
\citep{Hocker:2007ht} was trained. The analysis was then optimized for discovery with respect to the Boosted Decision Tree score. The optimized value resulted in a final data sample of $85\%$ atmospheric neutrinos and $15\%$ miss-reconstructed cosmic ray muons in the off time data sample (any events not within $\pm$2 hours of a GRB). An unbinned maximum likelihood search \citep{ic22_grb} was performed and each event passing the boosted decision tree cut was assigned a probability of being a signal event from a GRB or a background event. 

\subsection{ Results from triggered offline analysis}
No events were found in the on-time data to be on-source (within $10^\circ$ of a GRB) and on time with a GRB in the IC59 data set. In total 24 background events (not necessarily on source) were expected to be in the total time window and 21 were observed (none on-source). From the \cite{guetta2004} model 5.8 signal events were predicted and a final upper limit of 0.46 times the predicted flux can be set. This limit includes a $6\%$ systematic uncertainty. 
The corresponding model dependent result presented in a previous analysis \citep{Abbasi:2011qc} sets a limit of 0.82 of the model flux. 
Combining both results in a limit of 0.22 times the flux calculated according to \cite{guetta2004}.
\begin{figure}[t]
  \centering
  \includegraphics[width=\linewidth]{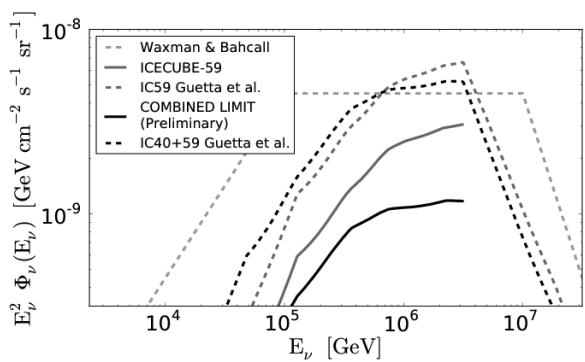}
  \caption{The combined limit of the IC40+59 analysis is shown in addition to the flux predictions from \cite{guetta2004} and \cite{Waxman2003}.}
  \label{fig:combined}
\end{figure}
This result implies that the model in question is strongly disfavored. 
The caveat is that certain parameters in the model, such as the bulk Lorentz factor $\Gamma$, are not (or only rarely) measured and therefore theoretically calculated values with large uncertainties are used. 
A lower limit on the Lorentz factor is established by pair production arguments \citep{guetta2004}, but the upper limit is less clear. Recent papers \citep{Bregeon:2011bu,Liang:2009zi,Soderberg:2002yr} 
suggest that $\Gamma$ can take values of up to 1000 (316 was used in this analysis), but it is questionable whether this is valid for a large fraction of GRBs. Whether the uncertainties on other parameters can account for the strict upper limit is currently being investigated.

\section{Online Analysis - Optical Follow-Up Program}
The online analysis triggers a network of optical telescopes by the detection of neutrino bursts (at least two neutrinos in a $100$\,s time window and an angular difference between their reconstructed directions of less than $4^{\circ}$).

\subsection{Online Event Selection}
\label{subsection:eventSelection}
In order to rapidly trigger optical telescopes the first online analysis of high-energy neutrinos detected by IceCube was developed and implemented.
The online analysis presented here was designed to be sensitive to neutrino production in soft relativistic SN jets. Straight cuts are applied in order to separate signal from background events.
Restricting the search to the Northern hemisphere and imposing requirements on the event reconstruction quality (e.g. the number of hits with small time residual or the likelihood of the reconstruction) allows a suppression of the mis-reconstructed muon background. 
To suppress the background of atmospheric neutrinos, which we cannot distinguish from the soft
SN neutrino spectrum, we require the detection of at least two events within 100\,s and an angular
difference between their two reconstructed directions of $\Delta\Psi \leq 4^{\circ}$. The choice of the time window
size is motivated by the observed $\gamma$-ray
emission from long GRBs (typically $50$\,s)
, which roughly corresponds to the time for a highly relativistic jet to penetrate the stellar envelope.
The angular window $\Delta\Psi$ is determined by the angular resolution of IceCube and was optimized along with the other selection parameters.
The final set of selection cuts has been optimized in order to reach a multiplet rate of $\sim25$ per year corresponding to the maximal number of alerts accepted by ROTSE. The final data stream consists of 37\% (70\%) atmospheric neutrinos for IC40 (IC59). Combining the neutrino measurement with the optical measurement allows the cuts 
to be relaxed yielding a larger background contamination and at the same time a higher signal passing rate. A doublet is not significant by itself, but may become significant when the optical information is added.
The resolution of the doublet direction is $\sim 0.8^{\circ}$.

\subsection{Search for Optical Counterparts}
\label{sec:OpticalCounterpart}
The IceCube multiplet alerts are forwarded to
the robotic optical transient search experiment (ROTSE), which consists of four 
identical telescopes each with a 45\,cm mirrow located in Australia, Texas, Namibia and Turkey \citep{ROTSE}. 
The telescopes stand out because of their large field of view of $1.85^{\circ} \times 1.85^{\circ}$ 
and a rapid response. 
Once an IceCube alert is received by one of the telescopes,
the corresponding region of the night sky will
be observed within seconds. Follow-up observations are performed regularly for 3 weeks. 
In the initial phase with IC40 and IC59, the online processing latency of several hours made the search for an optical GRB afterglow unfeasible. We therefore focus on the SN light curve detection in the ROTSE data.\\
Image subtraction followed by a detection algorithm to identify transient source candidates are applied to find an optical counterpart.

\begin{figure}[t!]
\centering
\includegraphics[angle=0,width=9cm]{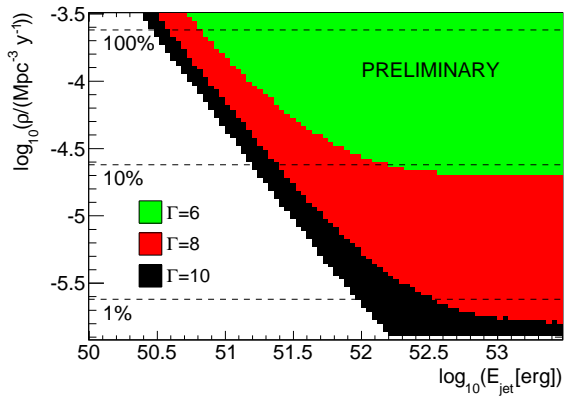}
\caption{Limits on the choked jet SN model \citep{AndoBeacom} for different boost Lorentz factors $\Gamma$ as a function of the rate of SNe with jets $\rho$ and the jet energy $E_{\rm{jet}}$ (colored regions are excluded at 90\% CL). 
Horizontal dashed lines
indicate a fraction of SNe with jets of 100\%, 10\% or 1\%.}
\label{pic:limit}
\end{figure}

\subsection{Results of the Optical Follow-up Program}
\label{sec:Results}

Here the results from the analysis of data taking in the period of 2008/12/16 to 2009/12/31 are presented. 
Table~\ref{tab:multiplets} shows the number of detected and expected doublets and triplets for the IC40 and the IC59 datasets as well as the number of detected and expected optical SN counterparts.
The IceCube expectation based on a background only hypothesis
was obtained from scrambled datasets. 
The number of doublets shows a small excess, which corresponds to a 2.1\,$\sigma$ effect and is thus not statistically significant.
\begin{table}[b]
 \caption[]{Measured and Expected Multiplets}
  \label{tab:multiplets}
 \begin{center}
  \begin{tabular}{| l | c | c  c | c  c |}
    \hline
                  & SN      & \multicolumn{2}{c|}{Doublets}     & \multicolumn{2}{c|}{Triplets}   \\
                  &         & IC40 & IC59                      & IC40 & IC59 \\
    \hline
    measured      & 0       & 15   & 19                        &   0 & 0 \\
    \hline
    expected      &  0.074 & 8.55 & 15.66                     &  0.0028 & 0.0040 \\
    \hline
  \end{tabular}
 \end{center}
\end{table}
The expected number of randomly coincident SN detections, $N_{\rm{SN}}^{bg}=0.074$, 
is based on an assumed core-collapse SN rate 
of 1 per year within a sphere of radius 10\,Mpc, i.e.\
$2.4\cdot10^{-4}$\,y$^{-1}$\,Mpc$^{-3}$, and a Gaussian absolute magnitude distribution
with mean of $-18$\,mag and standard deviation of $1$\,mag for CCSN \citep{SNMag}. 
No optical SN counterpart was found in the data.\\
We obtain the confidence level for different combinations of SN model parameters \citep{AndoBeacom} by using a pre-defined test statistic based on a likelihood function.
The limit is calculated for the jet boost Lorentz factors $\Gamma = 6,8,10$ as a function of the rate of SNe with jets $\rho$ 
and the jet energy $E_{\text{jet}}$. Systematic errors related to the simulated neutrino sensitivity and the SN sensitivity are included in the limit calculation. The 90\% confidence regions for each $\Gamma$-value are displayed in the $E_{\text{jet}}$-$\rho$-plane in Fig.~\ref{pic:limit}. 
The most stringent limit can be set for high $\Gamma$-factors. 
A sub-population fo SN with typical values of $\Gamma=10$ and $E_{\rm{jet}}=3\cdot10^{51}$\,erg does not exceed $4.2$\% (at $90$\% confidence). This is the first limit on CCSN jets using neutrino information.\\
Because of the successful operation of the optical follow-up program with ROTSE, the program was extended
in August 2010 to the Palomar Transient Factory \citep{PTF2}, which will provide deeper images and a fast
processing pipeline including a spectroscopic follow-up of interesting SN candidates. Furthermore, an X-ray follow-up by the Swift satellite of the most significant multiplets has been set up and started operations in February 2011.

\bibliography{fermiBib}





\end{document}